# The Poker-Litigation Game

F.E. Guerra-Pujol

*University of Central Florida*
*College of Business Administration*
*Dixon School of Accounting*
*4000 Central Florida Blvd.*
*Orlando, Fla. 32816, USA*

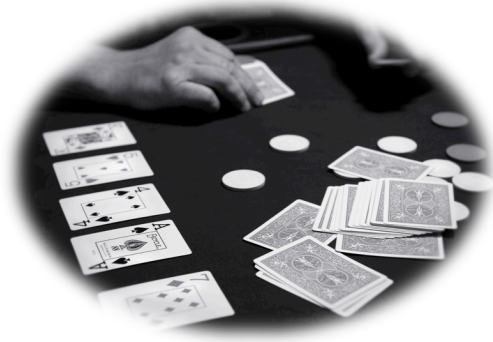

*Abstract*: Is litigation a search for truth, like science or philosophy, or a game of skill and luck, like the game of poker? Although the process of litigation has been modeled as a Prisoner's Dilemma, as a War of Attrition, as a Game of Chicken and even as a simple coin toss, no one has formally modeled litigation as a game of poker. This paper is the first to do so. We present a simple "poker-litigation game" and find the optimal strategy for playing this game.

# The Poker-Litigation Game

F.E. Guerra-Pujol[1]

*There is an undeniable . . . symmetry between law practice and poker . . . .* [2]

## 1. Introduction

Is litigation a search for truth, like science or philosophy, or a game of skill and luck, like the game of poker?[3] If the former, then how do we explain the occurrence of frivolous claims or negative-expected value lawsuits? If the latter, how do we explain the demand for costly methods of dispute resolution and the critique of random methods of justice?

Although the process of litigation has been modeled as a Prisoner's Dilemma,[4] as a War of Attrition,[5] as a Game of Chicken,[6] and even as a simple coin toss,[7] and although, in the words of one scholar, "there is an undeniable . . . symmetry between law practice and poker,"[8] to our knowledge no one has formally modeled litigation and the legal process as a game of poker. This paper is the first to do so. Specifically, we present a simple "poker-litigation game" and then find the optimal strategy for playing this game.

The remainder of this paper is organized as follows: Section 2 summarizes the similarities between litigation and the game of poker. Next, Section 3 presents a simple model of our poker-litigation game, and then Section 4 presents a formal solution of the game based on the work of David L. McAdams.[9] Section 5 concludes.

---

[1] B.A., U.C. Santa Barbara. J.D., Yale Law School, Lecturer, University of Central Florida (UCF), College of Business Administration, Dixon School of Accounting.
[2] *See* Steven Lubet, *Sidebar: The Game is Lawyer's Poker*, 32 LITIG. 59, 59 (2005).
[3] For an overview of the game of poker, *see* JOHN SCARNE, SCARNE ON CARDS 224-40 (rev. aug. ed. 1965).
[4] Ronald J.Gilson & Robert H. Mnookin, *Disputing Through Agents: Cooperation and Conflict Between Lawyers in Litigation*, 94 COLUM. L. REV. 509, 514-22 (1994).
[5] Paul Klemperer, *Why Every Economist Should Learn Some Auction Theory*, SOCIAL SCIENCE RESEARCH NETWORK (Oct. 12, 2000), available at http://papers.ssrn.com/sol3/papers.cfm?abstract_id=241350.
[6] F.E. Guerra-Pujol, *Coase and the Constitution*, 14 RICH. J.L. & PUB. INT. 593, 595-97 (2010).
[7] F.E. Guerra-Pujol, *Chance and Litigation*, 21 B.U. PUB. INT. L.J. 45, 46-48 (2011).
[8] Lubet, *supra* note 2, at 59-70.
[9] *See* David L. McAdams, *World's Simplest Poker*, CHEAP TALK (Nov. 20, 2012), available at http://cheeptalk.files.wordpress.com/2012/11/worlds-simplest-poker.pdf.

## 2. Similarities between poker and litigation

Litigation has many important features in common with the game of poker, for example:

(i) both poker and litigation are *strategic* games in which the players/litigants must make their choices independently of each other;[10]
(ii) both poker and litigation are *zero-sum, non-cooperative* games in which the economic interests of the players/litigants are opposed;[11]
(iii) both are games of *incomplete information*: just as a player in a game of poker does not know with certainty when another player is "bluffing," a litigant in a civil or criminal case may not know with certainty the strength of his adversary's case during pre-trial negotiations;[12]
(iv) both games involve significant elements of *chance* or luck: e.g. random assignment of the cards in poker; random selection of judges and jurors in civil and criminal cases.[13]

In addition, poker has a rich history of study in other academic fields, including mathematics,[14] game theory,[15] computer science,[16] and law.[17] This paper, however, is the first to formally model litigation as a game of poker. We present a simple model of our poker-litigation game in section 3 below and then find the optimal strategy for playing this game in section 4.

## 3. The model

Building on the work of John von Neumann, Oskar Morgenstern, John Nash, and others, we model litigation and the legal process as a game of poker.[18] Specifically, our poker-litigation game proceeds in four successive stages as follows:

(i) time $T_1$ . . . an opening round in which each player is dealt one card

---

[10] *See, e.g.*, John Nash, *Non-Cooperative Games*, 54 ANNALS MATH. 286, 286 (1951).

[11] *See, e.g.*, DOUGLAS G. BAIRD ET AL., GAME THEORY AND THE LAW 220-24 (1994).

[12] *See, e.g.*, STEVEN LUBET, LAWYER'S POKER: 52 LESSONS THAT LAWYERS CAN LEARN FROM CARD PLAYERS 92-93 (2006).

[13] *See generally* NEIL DUXBURY, RANDOM JUSTICE: ON LOTTERIES AND LEGAL DECISION-MAKING (1999).

[14] Nash, *supra* note 10, at 293-94; Harold W. Kuhn, *A Simplified Two-Person Poker*, *in* 1 CONTRIBUTIONS TO THE THEORY OF GAMES (Harold W. Kuhn & Albert W. Tucker, eds., 1950).

[15] JOHN VON NEUMANN & OSKAR MORGENSTERN, THEORY OF GAMES AND ECONOMIC BEHAVIOR (3d ed. 1953).

[16] Darse Billings, et al., *The Challenge of Poker*, ARTIFICIAL INTELLIGENCE 134 (2002).

[17] LUBET, supra note 12; *see, e.g.*, Scarne, *supra* note 3, in CHEATING AT BLACK JACK, for a historical overview and general description of poker.

[18] *See* Billings, *supra* note 16, at 237-39 for a glossary of poker terms and phrases. *See also* Scarne, *supra* note 3, at 228-32.

  (ii)  time $T_2$ . . . a quiet round in which the players examine their hole cards
  (iii)  time $T_3$ . . . a betting round in which the players place their bets
  (iv)  time $T_4$ . . . a final round in which the bets are paid to the winner

The rules and payoff structure of our poker-litigation game are as follows:

1. There are three players: (i) player P, the plaintiff, (ii) player D, the defendant, and (iii) player J, the dealer/judge. The objective of players P and D is to win the poker-litigation game (by maximizing their payoffs from the game), while the objective of the dealer/judge, by contrast, is to shuffle and deal the cards, collect and pay out the bets, and enforce and administer the rules of the game.[19]

2. Each player, P and D, is dealt a card "face down" by the dealer/judge at the start of play, i.e. time $T_1$. In summary, these "hole" cards (or private cards) represent the strength or weakness of each player's case. For simplicity and mathematical tractability, we assume that the values of the players' private cards are independent and identically distributed (i.i.d.) random variables on the interval [0, 1].[20] We further assume that cards with higher values (i.e. values nearer to 1) are deemed stronger (i.e. are worth more) than cards with lower values (values approaching 0).[21]

3. The players may (but are not required to) examine their hole cards at time $T_2$, although the values of their cards are not revealed until after the betting round.[22] In other words, each player/litigant is able to determine the strength or weakness of his case but does not know the strength or weakness of the other litigant's case.

---

[19] Previous mathematical models of the game of poker take the role of the dealer/judge for granted. *See, e.g.*, McAdams, *supra* note 9; VON NEUMANN & MORGENSTERN, *supra* note 15, at 186-219; Nash, *supra* note 10, at 293-94; John Nash & Lloyd S. Shapley, *A Simple Three-Person Poker Game*, *in* 1 CONTRIBUTIONS TO THE THEORY OF GAMES 105-16 (Harold W. Kuhn & Albert W. Tucker eds., 1950); and Kuhn, *supra* note 14, at 93-103. In our model, by contrast, the judge is an essential player, since her role is to detect cheating and monitor compliance with the rules of the game.

[20] That is, in place of a standard, finite deck of playing cards with values ranging from 2, 3, 4, 5, 6, 7, 8, 9, 10, Jack, Queen, King, and Ace, we assume an infinite deck of cards with values randomly ranging from 0 to 1.

[21] McAdams, *supra* note 9. *See also* VON NEUMANN & MORGENSTERN, *supra* note 15, at 187-88 ("[E]ach player draws a number $s + 1, \ldots, S$ instead. The idea is that $s = S$ corresponds to the strongest possible hand, $s = S – 1$ to the second strongest hand, etc., and finally $s = 1$ to the weakest … Thus the game begins with two chance moves: The drawing of number $s$ for player 1 and the for player 2, which we denote $s_1$ and $s_2$.") (ellipsis in original).

[22] *See infra* Rule 4. Note that not even the judge is allowed to see the private hole cards of the players until after the betting round.

4. After examining their hole cards, the players must simultaneously and secretly place their bets at time $T_3$.[23] Specifically, all bets must be placed in separate sealed envelopes and handed over to the judge, who will then award the combined bets to the winning player at the end of play at time $T_4$.[24]

5. The players are allowed to make only two possible bets, a high bet, *a*, or a low bet, *b*, where $a > b > 0$, and all bets made at time $T_3$ are final.[25] Thus, there is only one round of betting, and the players are not allowed to "call" or "re-raise."[26] Also, for further simplicity and tractability, we assume $a = 2b$.[27]

6. Lastly, the players reveal their cards at time $T_4$, and the dealer/judge declares a winner based on the following sub-rules:

    a. if both players P and D have submitted <u>high</u> bets, the player with the highest card wins both bets for a net gain of +*a* (in the event of a tie, the players get back their original bets and play again);

    b. if both players have submitted <u>low</u> bets, the player with the highest card wins both bets for a net gain of +*b* (in the event of a tie, the players get back their original bets and play again);

    c. if one player has submitted a high bet and the other a low bet, the player submitting the high bet automatically (by default) wins both bets for a net gain of +*b*, regardless of the values of the players' cards.

---

[23] *Cf.* VON NEUMANN & MORGENSTERN, *supra* note 15, at 188 ("The next phase of the general game of Poker consists of the making of 'Bids' by the players."). In this paper, we shall use the more common term "bets" instead of "bids."

[24] *See infra* Rule 6. As an aside, the bets of the players in this simple game can be likened to "investment levels" of the litigants. That is, during the process of the litigation, each litigant must decide (independently of the decision of the other litigant) how much effort to invest in his case. Unlike litigation, however, where the level of investment arguably has some effect on the outcome of the case, in poker the size of one's bet has no effect on the strength of one's hole cards.

[25] *Cf.* VON NEUMANN & MORGENSTERN, *supra* note 15, at 189-90 ("We shall express ... restrictions on the size of bids ... in the simplest possible form: We shall assume that the two numbers *a*, *b*

$$a > b > 0$$

are given ab initio, and that for every bid there are only two possibilities: the bid may be 'high,' in which case it is *a*; or 'low,' in which case it is *b*. By varying the ratio $a/b$ —which is clearly the only thing that matters—we can make the game risky when $a/b$ is much greater than 1, or relatively safe when $a/b$ is only a little greater than 1.").

[26] *Id.* at 186-88. We make these assumptions for mathematical tractability and ease of exposition.

[27] A high bet is like a "double-or-nothing" bet, since a high bet is twice as large as a low bet.

Although our model of the poker-litigation game is simple and artificial, it isolates two important variables of the game: (i) the size of each player's bet or level of investment in his case (the variables *a* and *b*), and (ii) the relative strength of his case (depending on his randomly-assigned hole card). Moreover, our model captures several essential features that poker and litigation have in common: both activities are strategic, zero-sum games of incomplete information involving elements of luck or chance. Specifically, the poker-litigation game is a strategic game because the players must choose their strategies (i.e. place their bets) independently, without communicating with each other, and their choices, once made, are final.[28] In addition, the poker-litigation game is a game of incomplete information, and this condition makes the solution of the game non-trivial. If the values of the players' hole cards were common knowledge (that is, if the players could see each other's cards before placing their bets), the solution would be trivial. The players would always make optimal bets: bet low when the value of one's card is lower than the other player's card to minimize one's losses.[29]

Given this simple set of rules (see Rules #1-6 above) and given the temporal and strategic structure of the game, what is the optimal or best strategy in the poker-litigation game?[30] Put another way, given that the poker-litigation game is a strategic game of incomplete information, when should a player bet high or bet low in order to minimize his losses and maximize his gains? Stated formally, does any strategy in this game guarantee a player a non-negative expected payoff?[31] Below, we proceed to find the solution or symmetric equilibrium (in pure or mixed strategies) of the poker-litigation game.

## 4. Solution

The most well-known solution concept in game theory is the Nash equilibrium.[32] Stated formally, a game has an equilibrium point when "no player can profitably deviate [i.e. play a different strategy], given the actions of the other players."[33]

---

[28] Nash, *supra* note 10, at 286 ("[E]ach participant acts independently, without collaboration or communication with any of the others."); *see also* MARTIN J. OSBORNE & ARIEL RUBINSTEIN, A COURSE IN GAME THEORY 14 (1994) ("For a situation to be modeled as a strategic game . . . the players [must] make decisions independently, no player being informed of the choice of any other player prior to making his own decision.").

[29] *See* Alvin E. Roth & Michael Malouf, *Game-Theoretic Models and the Role of Information in Bargaining*, 86 PSYCHOLOGICAL REVIEW 574 (providing an overview of the role of information in game theory); *see also* BAIRD ET. AL., *supra* note 10, at 79-121.

[30] Stated formally, what is the equilibrium strategy or Nash equilibrium of players P and D?

[31] *Cf.* McAdams, *supra* note 9.

[32] Nash, *supra* note 10, at 286; *see, e.g.*, OSBORN & RUBINSTEIN, *supra* note 28 at 14-15; *see also* BAIRD, ET AL., *supra* note 11 at 19-23.

[33] OSBORNE AND RUBINSTEIN, *supra* note 28 at 15.

Stated simply, a strategy is a Nash equilibrium when it is a best response given the possible choices of the other players.

Following the work of David McAdams,[34] who applies Nash's contradiction method of analysis,[35] we conjecture that an equilibrium exists with a threshold $t$ and probability $p$ such that a player always bets high given any card whose value is greater than $t$ (i.e. when such player has a "strong" case) and bets low with probability $p$ given any card whose value is lesser than $t$ (when such player has a "weak" case).

To test this conjecture, we consider three different treatments or litigation scenarios: (i) case #1, when the value of a player's private card is equal to $t$ (in other words, when the player has a "marginal" or close case in which he is just as likely to win as to lose); (ii) case #2, when the value of his hole card is below $t$ by some unknown quantity $x$, or $t - x$ (i.e. when the player has a "weak" case); and (iii) case #3 when the value of his hole card exceeds the threshold $t$ by $x$, or $t + x$ (the player has a "strong" case).

4.1. case #1: *t*

Assume that a given player (say, player P, the plaintiff) likes to hedge his bets: he bets high, $a$, when the value of his hole card is greater than or equal to a certain threshold, $t$, and bets low, $b$, when his card is below this threshold.[36] In plain English, the intuition behind this strategy is that one should place larger bets the stronger one's card is in order to maximize one's gains, since one has a higher chance of winning the game when one has a strong card.[37]

Now, assume that the value of Player P's card is equal to $t$ and, in addition, assume that his opponent, player D, also likes to hedge his bets.[38]

Player P must consider two possible scenarios: either player D's card is greater than $t$ (in which case player D bets high, $a$) or it is lesser than $t$ (in which case D bets low, $b$). If player D bets low $b$ because his hole card is lesser than $t$, then player P will win $+b$ regardless whether he, player P, makes a high or low bet. By contrast, if the value of player D's card is greater than $t$, then player P will lose his bet (regardless whether he, player P, has made a high or low bet), but because player P loses more in this scenario when he bets high than when he bets low, player P should prefer to bet low in this case in order to minimize his losses, since $b < a$. But notice that this preference is inconsistent with player P's strategy of making high bets when the value of his hole card is greater than or equal to the threshold $t$; therefore, a mixed or $m$-type strategy cannot be an equilibrium strategy or best response for player P.

---

[34] *See* McAdams, *supra* note 9.
[35] Nash, *supra* note 10, at 293.
[36] For reference, we shall designate this strategy as an "*m*-type" strategy or mixed strategy.
[37] And conversely, one should make smaller bets—the weaker one's card—in order to minimize one's losses.
[38] Stated formally, assume player D is also an *m*-type player. That is, assume player D is playing the same *m*-type strategy against player P that P is playing against D.

This analysis leads us to a larger point about our game: a player can still lose even with a high-value hole card, depending on the hole card of the other player, since the ultimate outcome of the game depends on the respective values of the hole cards of both players (and conversely, a player can still win even when he holds a low-value hole card). In other words, like litigation, playing this game and making bets are risky activities because there are costs and benefits of making high bets (and low bets). In equilibrium, these costs should be equal to the benefits—thus we proceed to define these costs and benefits to find the equilibrium strategy of this game.

First, suppose that the value of player P's card is equal to $t$ and that the other player's card (player D's card) is greater than $t$, an event which occurs with probability $1 - t$.[39] Player P will thus lose the game regardless whether he bets high or bets low, but he loses $-a$ when he bets high and $-b$ when he bets low. (That is, in this case player P loses an additional amount, the difference between $a$ and $b$, and this differential loss is equal to $b$ since we have previously assumed that $a = 2b$.) Thus, from player P's perspective, the true cost a making a high bet in this scenario is $b(1 - t)$.

Next, suppose player D's private card is below the threshold $t$, an event which occurs with probability $t$.[40] Also, suppose player D bets high in this scenario with probability $p$. When player D bets low in this case, player P wins $+b$ (the value of player D's low bet) regardless whether player P himself bets high or low. But when player D bets high (an event which occurs with probability $p$), then player P wins $+a$ (or $2b$ since $a = 2b$).[41]

Thus, the benefit to player P from making a high bet in this scenario (i.e. when player D's card is below $t$) can be stated formally as follows:

$(b + a)(t)(p)$, or $3b \times t \times p$ (since $a = 2b$), or $3btp$.

---

[39] *See supra* Rule #2. This event occurs with probability $1 - t$ because we are assuming that the possible values of each player's card in this game are independent and identically distributed (i.i.d.) random variables on the interval [0, 1]. Thus, if $t$ is the probability that player P or D's card is equal to $t$, then $1 - t$ is the probability that player D's card is greater than $t$.

[40] *See supra* note 39. This event occurs with probability $t$ because, by definition, if player D's card is above $t$ with probability $1 - t$, then his card will be below $t$ with probability $t$.

[41] As an aside, it is worth asking, why would player D ever bet high with a low-value card? Because if player P placed a low bet instead of a high one, then player D would have won $+b$ instead of losing $-a$.

In equilibrium, the costs and benefits of making high bets in both scenarios must be the same:

$b(1 - t) = 3btp$, or equivalently (after simplification): $1/t = 1 + 3p$

Furthermore, in equilibrium, this equality must hold not only at the threshold $t$ but also everywhere above and below $t$. Thus, we consider two additional scenarios or cases: case #2 in subsection 4.2 below, when the value of a player's hole card is below $t$ (i.e. when such player has a "weak" case), and case #3 in subsection 4.3, when his card exceeds $t$ (when he has a "strong" case).

4.2. case #2: $t - x$

We now proceed to determine a player's best strategy or best response when the value of his hole card is less than $t$.

For simplicity, consider this second type of case from player P's perspective.[42] Assume the value of player P's private card is below the threshold $t$ by some unknown quantity $x$, or $t - x$. In summary, player P has two options in his strategy set: he can either make a high bet, $a$, or make a low bet, $b$. But what are player P's payoffs for each such strategy in this case compared to his payoffs in the first type of case (i.e. when the value of player P's private card is equal to $t$)?

In summary, if player P bets high in this second type of case (i.e. when the value of his hole card is $t - x$), then player P loses $-a$ rather than winning $+a$ only when two conditions are met: when (i) the value of player D's private hole card lies on the interval $[t - x, t]$ and (ii) player D bets high. Since player D will bet high in this particular scenario (i.e. when his hole card is on the interval $[t - x, t]$) with probability $p$ times $x$ (recall that by assumption a player bets high with probability $p$ when the value of his hole card is below $t$), player P's payoff in this scenario is $-2apx$ lower (or stated equivalently: $-4bpx$ lower, since $a = 2b$) than when the same scenario occurs in the first type of case (when the value of player P's hole card is equal to $t$).

If, however, player P bets low in this second type of case, then he loses $-b$ rather than winning $+b$ only when two conditions are met: (i) when the value of player D's card lies on the interval $[t - x, t]$, and (ii) when player D bets low. Since player D will bet low with probability $(1 - p)x$, player P's payoff is now $-2b(1 - p)x$ lower in this scenario than when the same scenario occurs in the first type of case (threshold = $t$).

---

[42] Recall, however, that our analysis also applies to player D since the payoffs in this game are symmetrical.

Next, to find player P's best response, we set player P's revised payoffs for both strategies equal to each other and simplify as follows:[43]

$2p = (1 - p)$, or $p = 1/3$

In plain words, randomizing between high and low bets is a best response given any card less than *t*. Or, stated formally, a player (player P and, by symmetry, player D) is indifferent between making high or low bets for all values of *x* only when the other player is placing high bets with probability 1/3. Otherwise, if one of the players were placing high bets with a probability greater than or lesser than 1/3, the other player could adjust his betting strategy accordingly to increase his gains (or reduce his loses). Furthermore, given that $p = 1/3$, when we substitute this value for *p* in our original equilibrium equation, $1/t = 1 + 3p$, and solve for *t*, we see that $t = 1/2$ or 0.5. Thus, the optimal threshold is 0.5 and a player should bet high with probability 1/3 (or bet low with probability 2/3) when the value of his hole card is below this critical threshold.

4.3. case #3: *t + x*

Lastly, we wish to confirm a player's best strategy (best response) when the value of his hole card is greater than *t*.

Again, for simplicity, consider this third type of case from a particular player's perspective, player P, although the same analysis also applies to player D since this game is symmetrical, but now, assume the value of player P's private card exceeds the threshold *t* by some unknown quantity *x*, or *t + x*. As before, player P has two possible moves in this game (i.e. he can make a high bet *a* or a low bet *b*), so given this case (i.e. when the value of player D's card lies on the interval [*t*, *t + x*]), we proceed to find player P's payoffs for each such strategy compared to his payoffs in the first type of case (when the threshold is set to *t*).

If player P makes a high bet in this case (i.e., when his card exceeds *t* by an amount *x*), then he wins +*a* rather than losing –*a* when the value of player D's card lies on the interval [*t*, *t + x*], since by assumption a player always makes a high bet a when his card is greater than *t*. Thus, player P's payoff in this case is +2*ax* higher (or, +4*bx* higher, since *a* = 2*b*) than his payoff when the same scenario occurs in the first type of case (when threshold = *t*).

If, however, player P bets low in this scenario, he wins +*b*, which is the same amount he would have won in this scenario in the first type of case (when threshold = *t*), when the same two conditions are met: when (i) the value of player D's hole card is less than *t* and (ii) player D makes a low bet. Since these are the same conditions under which player P wins in the first type of case, player

---

[43] We set these "*t – x*" payoffs equal to each other because, in equilibrium, a player (player P and, by symmetry, player D) should be indifferent between making high and low bets.

P's payoff in this third scenario (case #3) is identical to his payoff when this same scenario occurs in case #1.

The payoff from making a high bet is increasing over the interval $[t, 1]$, and the payoff from making a low bet is constant, since by definition $b = b$, player P (and, by symmetry, player D) strictly prefers to bet high when the value of his hole card is above $t$. Further, since this is a symmetrical game, this same conclusion applies to player D.

To recap, given the rules and payoff structure of our game, a player should bet high with probability 1/3 (or bet low with probability 2/3) when the value of his private card is below $t$ and bet high in all other cases.

## 5. Conclusion

Before concluding, we wish to acknowledge and say a few words about the artificiality of our poker-litigation game,[44] or in the words of Nate Silver, we wish to explain why our model is "sophisticatedly simple."[45] Admittedly, our poker-litigation game is a simple representation or model of a more complex activity (civil and criminal litigation), but parsimony and simplicity can also be a virtue for several reasons.[46] One is tractability. A simple model is easier to analyze and work with than the "real world" is. That is, a simpler representation of poker (or litigation) is more tractable than the actual real-world game being modeled, or in the words of John von Neumann and Oskar Morgenstern, the founders of game theory, "actual Poker [like actual litigation, we would add] is really much too complicated a subject for an exhaustive discussion."[47] Another virtue of simplicity is clarity. The process of creating a simple model requires us to identify the players and their strategy sets, clearly define our terms, and explicitly state our assumptions, and unlike a purely verbal description of reality, a formal model allows us to make falsifiable predictions about the real world. But most importantly, a simple well-designed model can capture the essence of a strategic interaction that is present in a more complex real-world situation. Specifically, our simple model of litigation can help us isolate and demonstrate some fundamental features of the legal process.

In this paper, we have presented a simple model of the process of litigation, the "poker-litigation game," based on the premise that litigation has much in common with the game of poker. In summary, our simple model is useful because it isolates two important variables of the game: (i) the size of each player's bet or level of investment in his case (captured by the variables *a* and *b*), and (ii) the relative strength of his case (captured by the variable *t*). In addition, our model provides an alternative explanation of the existence of frivolous claims (or "negative-expected value" lawsuits) as well as prosecutorial misconduct, for one of the main lessons of our model is that (from the perspective of the players) there is an optimal level of bluffing in the poker-litigation game, with bluffing defined as placing a high bet probabilistically or randomly even when the strength of one's case is weak.

---

[44] *See* VON NEUMANN & MORGENSTERN, *supra* note 15; *see also* Nash, *supra* note 10; *see also* Nash & Shapley, *supra* note 19; *see also* Kuhn, *supra* note 14 (all of whom invented simplified poker games to illuminate fundamental principles of game theory).

[45] NATE SILVER, THE SIGNAL AND THE NOISE: WHY SO MANY PREDICTIONS FAIL—BUT SOME DON'T 225 (2012).

[46] *See, e.g.*, VON NEUMANN & MORGENSTERN, *supra* note 15, at 186-88; *see also* F.E. Guerra-Pujol, *A Game-Theoretic Analysis of the Puerto Rico Status Debate and Other Legislative Wars of Attrition*, 18 AM. U. J. GENDER SOC. POL'Y & L. 625, 630-31 (2010).

[47] VON NEUMANN & MORGENSTERN, *supra* note 15, at 186.

But what is this optimal level of bluffing in our game? Given the rules and payoff structure of our game, the optimal level is to bet high with probability 1/3 when the value of one's private card is below the critical threshold 0.5. Likewise, in real-world litigation games, we would expect the optimal level of bluffing (i.e. frivolous claims) to be a function of two variables: (i) the amount at stake, or in the language of our model, the sum of the bets placed, and (ii) the relative strength and weakness of player P and D's cases.

## Acknowledgments

First and foremost, the author wishes to thank Jeffrey Ely for posting on cheaptalk.org David McAdams's poker game ("World's Simplest Poker") as well as McAdams's simple and elegant solution to the game. As an aside, all the solutions, save for one, posted in response to Professor Ely's initial blog post (including my own solution), were wrong. (See Appendix for my initial failed attempt to solve the game.) I must also thank professors Ely and McAdams for taking time during the holidays to answer my technical questions about the McAdams model. I developed my poker-litigation game and wrote up a first draft of this paper during the end of year holidays in 2012 at the home of my dear friends Renard and LaTanya Damon in Cooper City, Florida, at the home of my gracious in-laws, Erle and Andrea Robinson, in Tarpon Springs, Florida, and at the home of my caring and generous parents, Francisco and Oilda Guerra, in Glendale, California. Thus I also wish to thank the Guerras, the Robinsons, and the Damons for their kindness, support, and hospitality. Lastly, I wish to thank my wife Sydjia Guerra for proofreading and double-checking the mathematics in my paper. (The image on my cover page is courtesy of Wikimedia Commons and is available at http://commons.wikimedia.org/wiki/File:Holdem.jpg.)

## Bibliography

Baird, Douglas G., Robert H. Gertner, and Randal C. Picker (1994), *Game theory and the law*, Harvard.

Billings, Darse, Aaron Davidson, Jonathan Schaeffer, and Duane Szafron (2002), "The challenge of poker," *Artificial Intelligence*, 134:201-240.

Duxbury, Neil (1999), *Random justice*, Oxford.

Ely, Jeffrey (2012), "Great prelim question: world's simplest poker game," blog, Nov. 20, 2012, available online, URL = http://cheaptalk.org/2012/11/20/great-prelim-question-worlds-simplest-poker-game (last visited on Dec. 12, 2012).

Gilson, Ronald J., and Robert H. Mnookin (1994), "Disputing through agents: cooperation and conflict between lawyers in litigation," *Columbia Law Review*, 94(2):509-566.

Guerra-Pujol, F.E. (2010a), "Coase and the constitution," *Richmond Journal of Law and the Public Interest*, 14(4):593-609.

Guerra-Pujol, F.E. (2010b), "A game-theoretic analysis of the Puerto Rico status debate and other legislative wars of attrition," *American University Journal of Gender, Social Policy and the Law*, 18(3): 625-648.

Guerra-Pujol, F.E. (2011), "Chance and litigation," *Boston University Public Interest Law Journal*, 21(1):45-59.

Klemperer, Paul (2000), "Why every economist should learn some auction theory," invited paper for the World Congress of the Econometric Society held in Seattle in August 2000, available online, URL = http://papers.ssrn.com/sol3/papers.cfm?abstract_id=241350

(last visited on Dec. 12, 2012).

Kuhn, Harold W. (1950), "A simplified two-person poker," in Harold W. Kuhn and Albert W. Tucker, editors, *Contributions to the theory of games*, Princeton, vol. 1, pp. 97-103.

Lubet, Steven (2005), "Sidebar: the game is lawyer's poker," *Litigation*, 32(1):59-70.

Lubet, Steven (2006), *Lawyer's poker*, Oxford.

McAdams, David L. (2012), "World's simplest poker," unpublished paper, available online, URL = http://cheeptalk.files.wordpress.com/2012/11/worlds-simplest-poker.pdf (last visited on Dec. 12, 2012).

Nash, John, and Lloyd S. Shapley (1950), "A simple three-person poker game," in Harold W. Kuhn and Albert W. Tucker, editors, *Contributions to the theory of games*, Princeton, vol. 1, pp. 105-116.

Nash, John (1951), "Non-cooperative games," *Annals of Mathematics*, 54(2):286-295.

Osborne, Martin J., and Ariel Rubinstein (1994), *A course in game theory*, MIT.

Roth, Alvin E., and Michael W.K. Malouf (1979), "Game-theoretic models and the role of information in bargaining," *Psychological Review*, 86(6):574-594.

Roth, Alvin E., and J. Keith Murnighan (1982), "The role of information in bargaining: an experimental study," *Econometrica*, 50(5):1123-1124.

Scarne, John (1965), *Scarne on cards*, revised, augmented edition, Crown.

Silver, Nate (2012), *The signal and the noise: why so many predictions fail but some don't*, Penguin.

Von Neumann, John, and Oskar Morgenstern (1953), *Theory of games and economic behavior*, 3rd edition, Princeton.

## Appendix

As an aside, I set forth in this appendix my initial (and failed) attempt to solve the poker-litigation game (i.e. to find the optimal or best strategy in this game). I include my false start to contrast it with the correct solution to the game, which I obtained from David McAdams and Jeff Ely.

In summary, in my initial failed attempt to solve the game, I considered three types of players: "*a*-type" players who always make high bets, "*b*-type" players who always make low bets, and "*m*-type" players whose bets depend on the value of their hole cards.

*a*-type players

First, I considered an "*a*-type" player who always submits a high bet, *a*. An *a*-type player will lose his bet only when two conditions are met: (*i*) when the other player has made a high bid, and (*ii*) when his card has a lower value than the card of the other player.

Otherwise, an *a*-type player always win (as per Rule #6c above), and thus, if such a player plays the litigation game an infinite number of times, his expected payoff is equal to the sum of $pa - (1 - p)(a)$ when the other player makes a high bid—where $p$ is the probability that his card is higher than the other player's card—, and $+a$ when the other player makes a low bid.

<u>*b*-type players</u>

Next, I considered a "*b*-type" player whose strategy is to <u>always</u> place a low bet, *b*. In summary, a *b*-type player wins the game only when two conditions are met: (*i*) when his card has a higher value than the card of the other player, and (*ii*) when the other player has submitted a low bid. Otherwise, when neither of these conditions are met, a *b*-type player will always lose. Thus, if a *b*-type player plays the litigation game an infinite number of times, his expected payoff is equal to the sum of $-b$ when the other player submits a high bid and $pb - (1 - p)(-b)$ when the other player submits a low bid, where $p$ is the probability that his card is higher than the other player's card.

<u>*m*-type players</u>

Lastly, I considered an *m*-type player who plays a <u>mixed</u> or probabilistic strategy, that is, a player who makes a low bet, *b*, when his card is below a certain threshold, namely 0.5, and who makes a high bet, *a*, when his card is above this critical threshold.[48] An *m*-type player has many ways of winning or losing the litigation game, depending on what strategy the other player is choosing and on the value of other player's card. Thus, we find an *m*-type player's expected payoff against *a*-type players, *b*-type players, and other *m*-type players as follows:

First, an *m*-type player's expected payoff against an *a*-type player, i.e. a player who always makes <u>high</u> bids, is equal to the following value:

$E(m|a) = q(-b) + (1 - q)[pa - (1 - p)(-a)]$

where $E(m|a)$ is the expected payoff of an *m*-type player against an *a*-type player, $q$ is the probability that the *m*-type player's card is below the critical threshold value 0.5 and where $p$ is the probability that his, the *m*-type player's, card is higher than the other player's card. In other words, when playing against a high-bet, *a*-type player, an *m*-type player loses his bet *b* when the value of his hole card is below the critical threshold, i.e., when $q < 0.5$, and wins $+a$ with probability $p$ but loses $-a$ with probability $1 - p$ when the value of his hole card exceeds the critical threshold, i.e. when $q > 0.5$.

Next, what happens when an *m*-type player plays against a *b*-type player, that is, a player who always makes <u>low</u> bets? In this case, the *m*-type player's expected payoff is equal to:

$E(m|b) = q[pb - (1 - p)(b)] + (1 - q)(b)$

where $E(m|b)$ is the expected payoff of an *m*-type player against an *b*-type player and $q$ is the probability that the *m*-type player's card is below the threshold value 0.5. In plain

---

[48] Such a mixed strategy is probabilistic in nature because the value of one's hole card is a uniform independent random number on the interval [0,1] as per Rule #2 of the poker-litigation game.

English, when playing against a low-bid, *b*-type player, an *m*-type player wins +*b* with probability *p* but loses his bet *b* with probability 1 – *p* when the value of his hole card is below the critical threshold, i.e. when $q < 0.5$, and he wins the bid +*b* when the value of his hole card exceeds the critical threshold, i.e. when $q > 0.5$.

Third and last, what happens when an *m*-type player plays against another *m*-type player? That is, what happens when his opponent also plays the same mixed or probabilistic strategy? Now, the *m*-player's expected payoff not only depends on *q*, the probability that the *m*-type player's card is below the critical threshold value 0.5; his expected payoff is also a function of *r*, the probability that the other player's card is below the threshold. Stated formally, when playing against another *m*-type player, an *m*-type player's expected payoff is equal to the following value:

$$E(m|m) = qr[pb - (1-p)(b)] + q(1-r)[-b] + (1-q)(r)[b] + (1-q)(1-r)[pa - (1-p)(-a)]$$

where $E(m|m)$ is the expected payoff of an *m*-type player against an another *m*-type player, *q* is the probability that the *m*-type player's hole card is below the critical threshold value 0.5, and *r* is the probability that the other player's hole card is below this threshold. Since this is such a lengthy equation, we shall break it down into its four constituent parts and explain the substance of each part in plain words as follows:

$$\underbrace{qr[pb - (1-p)(b)]}_{\text{scenario \#1}} + \underbrace{q(1-r)[-b]}_{\text{scenario \#2}} + \underbrace{(1-q)(r)[b]}_{\text{scenario \#3}} + \underbrace{(1-q)(1-r)[pa - (1-p)(-a)]}_{\text{scenario \#4}}$$

Notice that each part of this lengthy expected payoff equation corresponds to one of following four possible scenarios:

1. Scenario #1—both players' hole cards are below the critical threshold.
2. Scenario #2—the first *m*-type player's hole card is below the threshold; theother player's hole card is above the threshold.
3. Scenario #3—the first *m*-type player's hole card is above the threshold, while the other player's hole card falls below the threshold.
4. Scenario #4—both players' hole cards exceed the critical threshold.

Thus, depending on which scenario occurs, that is, depending on the values of the hold cards of the players, an *m*-type player will earn the following payoffs:

1. In scenario #1, an *m*-type player wins +*b* with probability *p* but loses his bet *b* with probability 1 – *p*.
2. In scenario #2, an *m*-type player loses –*b*.
3. In scenario #3, an *m*-type player wins +*b*.
4. And in scenario #4, an *m*-type player wins +*a* with probability *p* but loses –*a* with probability 1 – *p*.

To recap, in my initial failed attempt to solve the poker-litigation game, I considered three types of players (or three types of strategies)—"*a*-type" players who always bet high, "*b*-type" players who always bet low, and "*m*-type" players who bet high or low depending on the value of their hole card—and I also figured out the expected payoffs of the strategies of each type of player. But it was at this stage that I was "stumped," unable to determine which strategy is the optimal strategy (and thus I was unable to find what type of player earns the highest expected payoffs in this game).

In short, because of my inability to solve the game using traditional methods, I instead turned to David McAdams's elegant solution to steer me in the right direction.